\documentclass{patmorin}
\usepackage{amsfonts,amsopn,amsthm,graphicx}
\usepackage[mathlines]{lineno}

\newcommand{\Z}{\mathbb{Z}}
\newcommand{\etal}{\emph{et~al.}}
\DeclareMathOperator{\ch}{ch}
\DeclareMathOperator{\ind}{indeg}
\DeclareMathOperator{\outd}{outdeg}

\newtheorem{lem}{Lemma}

\newcommand{\seclabel}[1]{\label{sec:#1}}

\newcommand{\secref}[1]{\mbox{Section~\ref{sec:#1}}}

\newcommand{\figlabel}[1]{\label{fig:#1}}

\newcommand{\figref}[1]{\mbox{Figure~\ref{fig:#1}}}

\newcommand{\eqlabel}[1]{\label{eq:#1}}

\newcommand{\eqref}[1]{(\ref{eq:#1})}

\newtheorem{thm}{Theorem}
\newcommand{\thmref}[1]{Theorem~\ref{thm:#1}}
\newcommand{\thmlabel}[1]{\label{thm:#1}}
\newcommand{\lemlabel}[1]{\label{lem:#1}}
\newcommand{\lemref}[1]{Lemma~\ref{lem:#1}}

\begin{document}

\title{Notes on Large Angle Crossing Graphs}%

\author{%
Vida Dujmovi\'c\thanks{{Carleton University, Canada}, \texttt{vida@mcgill.ca}},\,
Joachim Gudmundsson\thanks{{NICTA Sydney, Australia}, \texttt{joachim.gudmundsson@nicta.com.au}},\,
Pat Morin\thanks{{Carleton University, Canada}, \texttt{morin@scs.carleton.ca}},\,
Thomas Wolle\thanks{{NICTA Sydney, Australia}, \texttt{thomas.wolle@nicta.com.au}}\,
}

\maketitle

\begin{abstract}
A geometric graph $G$ is an $\alpha$ angle crossing ($\alpha$AC) graph if
every pair of crossing edges in $G$ cross at an angle of at least
$\alpha$.  The concept of right angle crossing (RAC) graphs ($\alpha=\pi/2$)
was recently introduced by Didimo \etal\ \cite{del-dgrac-09}. It was
shown that any RAC graph with $n$ vertices has at most $4n-10$ edges
and that there are infinitely many values of $n$ for which there exists a RAC
graph with $n$ vertices and $4n-10$ edges.  In this paper, we give upper
and lower bounds for the number of edges in $\alpha$AC graphs for all
$0 < \alpha < \pi/2$.
\end{abstract}

\section{Introduction}

The problem of producing visually appealing graph drawings of relational data sets is a fundamental problem and has been studied extensively, see the books~\cite{bett-gd-99,jm-gds-03,kw-dgmm-01,nr-pgd-04}.
One measure of a graph drawing algorithm's quality is the number of edge crossings it draws~\cite{ew-ecdbg-94,jm-2lscm-97,kw-dgmm-01,n-ibosm-05}.
While some graphs cannot be drawn without edge crossings, some graphs can. These are called planar graphs.
According to this metric, ``good'' algorithms draw graphs with as few edge crossings as possible.
This intuition has some scientific validity:
experiments by Purchase \etal~\cite{p-eiv-00,pca-eeabg-02,wpcm-cmga-02}
have shown that the performance of humans in path tracing tasks is negatively correlated to the number of edge crossings and to the number of bends in the drawing.

However, recently Huang \etal\ \cite{h-uetig-07,h-etseg-08,hhe-eca-08} showed, through eye-tracking experiments, that crossings that occur at angles of greater than $70^\circ$ have little effect on a human's ability to interpret graphs.  Therefore, graph drawings with crossing are not bad, as long as the crossings occur with large angles between them. This motivated Didimo \etal\ \cite{del-dgrac-09} to introduce so-called right angle crossing (RAC) graphs.  A \emph{geometric graph}~\cite{ae-degg-89} is a graph $G=(V,E)$ such that the vertices are distinct points in $\mathcal{R}^2$ and edges are straight-line segments. A geometric graph $G$ is a RAC graph if any two crossing segments are orthogonal with each other.

In this paper we generalize the concept to $\alpha$ angle crossing ($\alpha$AC) graphs.  A geometric graph $G$ is an $\alpha$AC graph if every pair of crossing edges in $G$ cross at an angle of at least $\alpha$.  Clearly, $\alpha$AC graphs are more general than planar graphs and RAC graphs, but how much more so?  One measure of generality is the maximum number of edges such a graph can represent.  Euler's Formula implies that a planar graph with $n$ vertices has at most $3n-6$ edges.  How many edges can an $\alpha$AC graph have?

\subsection{Previous Work}
Didimo \etal\ \cite{del-dgrac-09} studied $\pi/2$-angle crossing graphs, called right angle
crossing (RAC) graphs, and showed that any RAC graph with $n$ vertices has at most $4n-10$ edges and
that there exist infinitely many values of $n$ for
which there exists a RAC graph with $n$ vertices and $4n-10$ edges.
Recently, Angelini \etal~\cite{acddfks-porac-09} considered some
special cases for drawing RAC graphs, for example, acyclic planar RAC
digraphs and upward RAC digraphs. They showed that there exist acyclic planar
digraphs not admitting any straight-line upward RAC drawing and that the
corresponding decision problem is NP-hard. They also gave a construction of
digraphs whose straight-line upward RAC drawings require exponential area.

For $\alpha > \pi/3$, an $\alpha$AC graph has no three edges that
mutually cross since, otherwise, one of the pairs of edges must
cross at an angle that is at most $\pi/3$.  Geometric graphs with no
three pairwise crossing edges are known as \emph{quasiplanar graphs}
\cite{aapps-qpgln-97}. Ackerman and Tardos \cite[Theorem~5]{at-mneqp-07} have shown that
any quasiplanar graph on $n$ vertices has at most $6.5n - 20$ edges.

For $\alpha > \pi/4$, an $\alpha$AC graph has no four pairwise crossing edges.
Ackerman~\cite{a-mnetg-09} has shown that any such graph has at most $36n - 72$ 
edges.  It remains an open problem whether, for any $k\ge 5$, a graph with no
$k$ pairwise crossing edges has a linear number of edges.  The best known
upper bound of $O(n\log n)$ on the number of edges in such a graph is due
to Valtr \cite[Theorem~3]{v-epct-99}.

A recent paper of Arikushi \etal\ \cite{afkmt-dgoc-10} considers a similar question. They 
consider a drawing of a graph $G$ on $n$ vertices in the plane where each edge is represented 
by a polygonal arc joining its two respective vertices.  They consider the class of graphs in 
which each edge has at most one bend or two bends, respectively, and any two edges can cross 
only at right angle. It is shown that the number of edges of such graphs is at most $6.5n$ and 
$74.2n$, respectively. 

\subsection{New Results}

The current paper gives upper and lower bounds on the number of edges in
$\alpha$AC graphs.  In \secref{uniform} we show that, for any $0< \alpha
<\pi/2$, the maximum number of edges in an $\alpha$AC graph is at most
$(\pi/\alpha)(3n-6)$.  In \secref{lower-bounds}, we give constructions that
essentially match this upper bound when $\alpha = \pi/t-\epsilon$, for
any integer $t\ge 2$ and any $\epsilon > 0$.  Finally, in~\secref{charging} we use
a charging argument similar to the one used by
Ackerman and Tardos~\cite{at-mneqp-07} to prove that, for
$2\pi/5 < \alpha < \pi/2$, the number of edges in an $\alpha$AC graph is
bounded by $6n-12$.
An overview of previous and new results is illustrated in \figref{fig:overview}.
\begin{figure*}[tbh]
  \begin{center}
    \includegraphics[width=14cm]{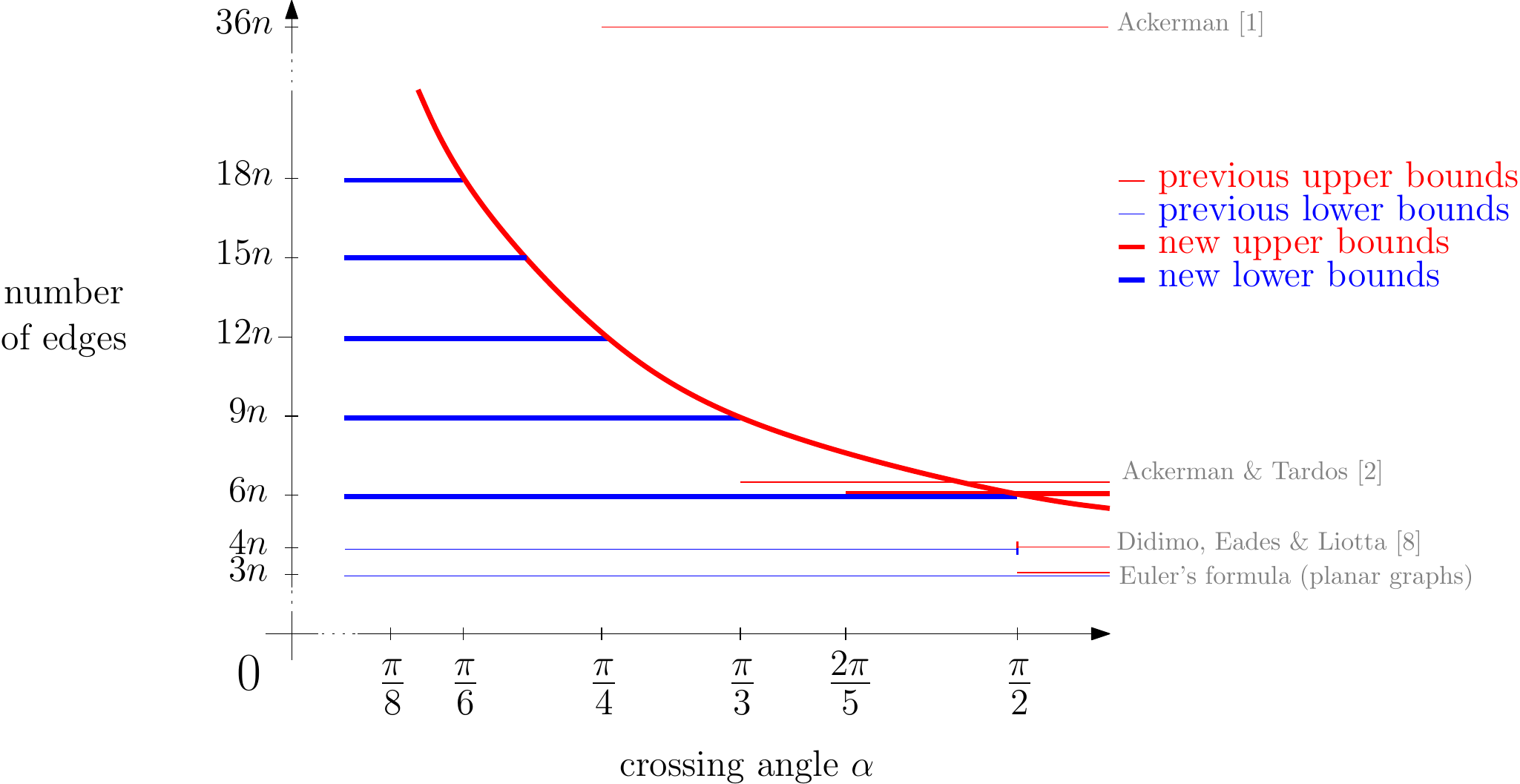}
  \end{center}
  \caption{A plot of previous and new upper and lower bounds. Lower order terms are disregarded.}
  \figlabel{fig:overview}
\end{figure*}

\section{A Uniform Upper Bound}
\seclabel{uniform}

In this section, we give an upper bound of $(\pi/\alpha)(3n-6)$ on the
number of edges in an $\alpha$AC graph.
This upper bound captures the intuition that an $\alpha$AC graph can be viewed as the union of $\pi/\alpha$
planar graphs. The only trouble with this intuition is that $\pi/\alpha$
is not necessarily an integer.

\begin{thm}\thmlabel{uniform-upper-bound}
Let $G$ be an $\alpha$AC graph with $n$ vertices for some $0<\alpha<\pi/2$.
Then $G$ has at most $(\pi/\alpha)(3n-6)$ edges.
\end{thm}

\begin{proof}
Define the \emph{direction} of an edge $xy$ whose lower endpoint is $x$ (in
the case of a horizontal edge, take $x$ as the left endpoint) as the angle 
$\angle wxy$ where $w=x+(1,0)$. The direction of an edge $xy$ is
therefore a real number in the interval $[0,\pi)$. Let $r=\lceil\pi/\alpha\rceil$. 
Now, take a random rotation $G'$ of $G$ and partition the edge set of $G'$ 
into spanning subgraphs $G_1,\ldots, G_{r}$, and $G_i$, $1\leq i\leq r$,
contains all edges of $G'$ whose direction is in the interval $[\alpha(i-1),\alpha i)$.

Note that no two edges of $G_i$ cross each other, so each $G_i$ is
a planar graph that, by Euler's Formula, has at most $3n-6$ edges.
Furthermore, since $G'$ is a random rotation, the expected number of
edges in $G_{r}$ is $(\pi\bmod \alpha)/\pi \cdot |E(G)|$.
In particular, there must exist some rotation $G'$ of $G$ such that $|E(G_r)| \le (\pi\bmod \alpha)/\pi \cdot |E(G)|$.
Therefore,
\begin{equation}
   E(G) \le \lfloor \pi/\alpha \rfloor(3n-6) + (\pi\bmod\alpha)/\pi \cdot|E(G)| \enspace .
   \eqlabel{uniform}
\end{equation}
Rearranging \eqref{uniform} yields
\[
  |E(G)|
    \le  \frac{\lfloor \pi/\alpha \rfloor(3n-6)}{1-(\pi\bmod\alpha)/\pi}
    = (\pi/\alpha)(3n-6) \enspace ,
\]
as required.
\end{proof}

\section{Lower Bounds}
\seclabel{lower-bounds}

In this section we give constructions that essentially match
\thmref{uniform-upper-bound} for any $\alpha$ of the form
$\alpha=\pi/t-\epsilon$, where $t\ge 2$ is an integer.  All our
lower-bounds are based on a general technique of defining a 3-dimensional
geometric graph and taking a 2-dimensional projection of this graph
to obtain an $\alpha$AC graph.  Before presenting the technique in
all its generality, we first present an \emph{ad hoc} construction
for $\alpha=\pi/2$ that still illustrates the main ingredients of the
technique.

\begin{thm} \label{thm:LB_thm2}
For any $\epsilon > 0$ and all sufficiently large integers $n$, there exist
$(\pi/2-\epsilon)AC$ graphs that have $n$ vertices and $6n- O(n^{2/3})$ edges.
\end{thm}

\begin{proof}
Consider the infinite planar graph whose vertices are the integer lattice
$\Z^2$ and in which each vertex $(i,j)$ is adjacent to $(i+1,j+1)$,
$(i+1,j+2)$ and $(i,j+1)$ (see \figref{infinite}.a).  This graph is
6-regular, and if we consider the subgraph of this graph induced by an
$x\times y$ grid, then this graph has $3xy - O(x+y)$ edges.  We call
this graph $G_{x\times y}$.

We will draw $2r$ copies of $G_{r\times r}$ in $\Z^3$ as follows.
For each $i\in\{0,\ldots,r-1\}$, we draw one copy on the ``vertical'' grid
$\{ (i,j,k) : j,k\in\{0,\ldots,r-1\}\}$ and one copy on the ``horizontal''
grid $\{ (j,i,k) : j,k\in\{0,\ldots,r-1\}\}$ (see
\figref{infinite}.b). Note that there are 4 possible ways of drawing $G_{r\times r}$ on these
$r\times r$ grids.  We choose one of the two ways that have the property
that no edge of $G_{r\times r}$ is drawn parallel to the $z$-axis.
The resulting 3-d geometric graph has $n=r^3$ vertices and $6r^3 -
O(r^2)=6n-O(n^{2/3})$ edges.  All that remains is to show how to
project this 3-d geometric graph to the plane without introducing small
crossing angles.

Consider the orthogonal projection of the above graph onto a plane whose
normal is only slightly skewed from the $z$-axis (\figref{infinite}.b
shows such a projection).  For example, we could choose a plane orthogonal
to the vector $(\gamma,\gamma,1)$ for some arbitrarily small $\gamma>0$.
Since our projection is not parallel to the $z$-axis, each individual copy of
$G_{r\times r}$ projects to a planar drawing of $G_{r\times r}$.  
Since our projection
is only slightly skewed, no edge of any horizontal (respectively,
vertical) graph intersects any edge of any other horizontal (respectively,
vertical) graph.  Finally, the projection of each edge in each horizontal
graph is a segment whose direction lies in the interval $[-\delta,\delta]$
and the direction of each edge in each vertical graph is a segment whose
direction lies in the interval $[\pi/2-\delta,\pi/2+\delta]$, where
$\delta>0$ can be made arbitrarily close to 0 by choosing a sufficiently
small value of $\gamma$.  Thus,
any pair of crossing edges cross at an angle of at least $\pi/2-2\delta \ge \pi/2-\epsilon$ for $\delta=\epsilon/2$.
\end{proof}

\begin{figure}
  \begin{center}
  \begin{tabular}{cc}
    \includegraphics[height=2in]{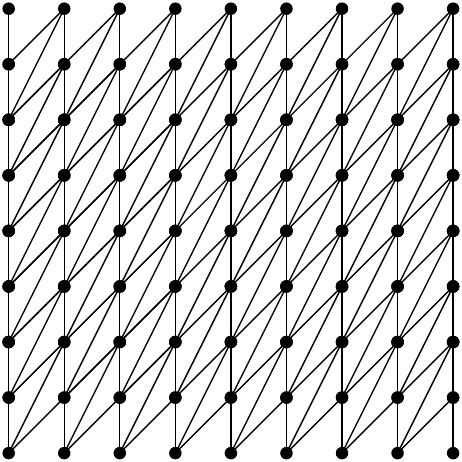} &
    \includegraphics[height=2in]{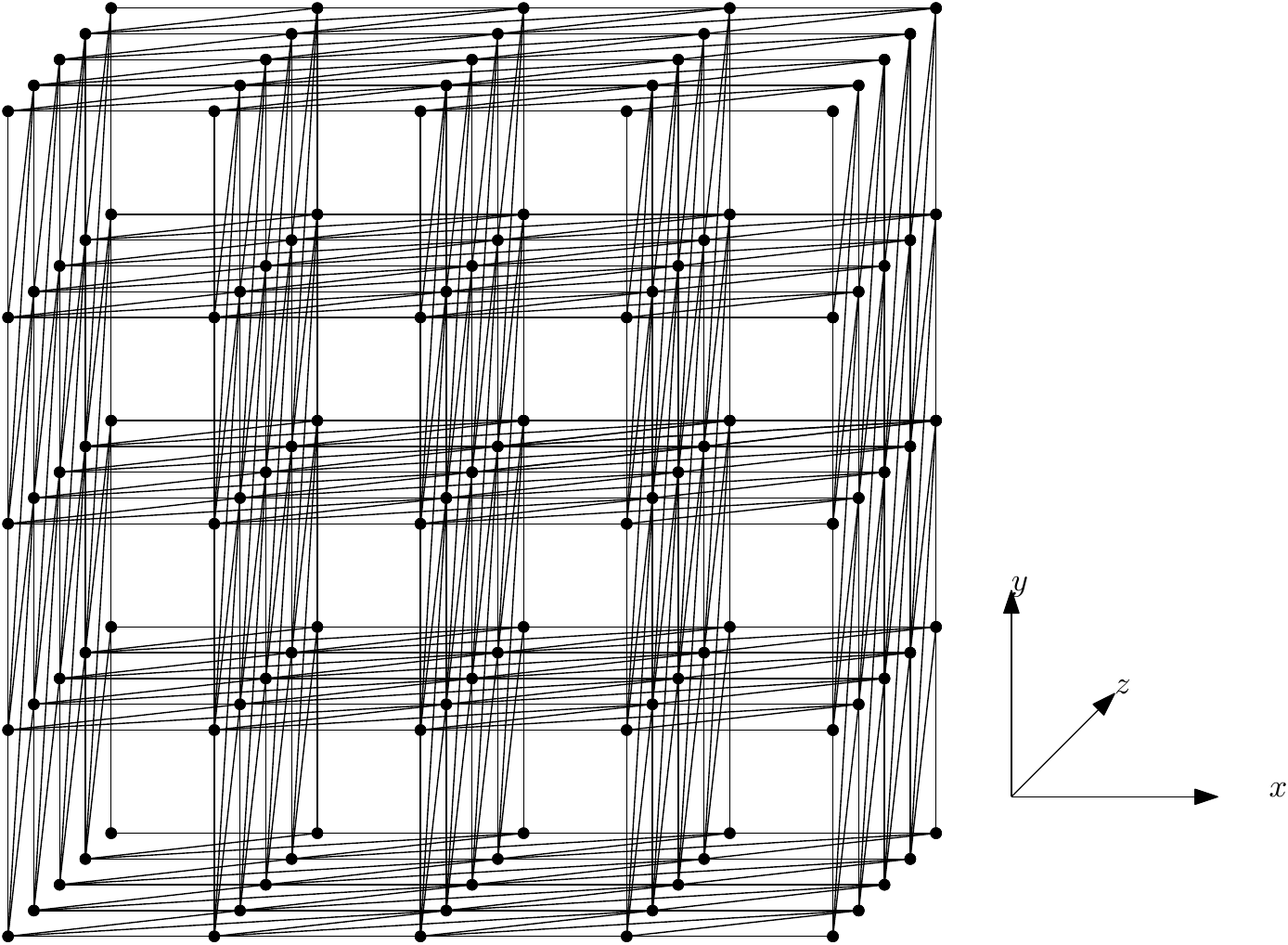} \\
    (a) & (b)
  \end{tabular}
  \end{center}
  \caption{The lower bound in \thmref{LB_thm2}.}
  \figlabel{infinite}
\end{figure}

\newcommand{\R}{\mathbb{R}}
The construction in the proof of \thmref{LB_thm2} can be viewed as first
starting with an arrangement of lines (in this case, $r$ horizontal
and $r$ vertical lines), extending each line into a plane in $\R^3$ and
drawing a copy of $G_{a\times b}$ on this plane so that the vertices lie
on the intersections of several planes.  The following lemma shows that
this can be done with any set of lines:

\begin{lem}\lemlabel{convert}
If there exists a set $S$ of $r$ lines in $\R^2$ such that
\begin{enumerate}
\item any two lines in $S$ are parallel or cross at an angle of at least $\alpha$ and
\item there are $x$ points in $\R^2$ each contained in at least $t$ lines
of $S$, 
\end{enumerate}
then, for any integer $k>1$ and any $\epsilon>0$, there exist
$(\alpha-\epsilon)$AC graphs that have $kx$ vertices and $3txk - O(kr+x)$
edges.
\end{lem}

For example, applying \lemref{convert} to a set of $r$ horizontal lines
and $r$ vertical lines with $k=r$ gives a $(\pi/2-\epsilon)$AC graph
with $n=r^3$ vertices and $6n - O(n^{2/3})$ edges, yielding the result
of \thmref{LB_thm2}.

\begin{proof}[Proof of \lemref{convert}]
Refer to \figref{convert} for an example.  As before, we create a 3-d
geometric graph.  Let $X$ be the set of points in $\R^2$ where $t$ or
more lines of $S$ intersect.  For each point $(x,y)\in X$, our graph
contains the vertices $\{(x,y,i) : i\in\{0,\ldots,k-1\}\}$.  Notice that
each line of $S$ that contains $p$ points in $X$ corresponds to a plane
in $\R^3$ that contains $pk$ points.  On each such plane, we draw a
copy of $G_{p\times k}$, creating $3pk-O(k+p)$ edges. Again,
we orient this drawing so that no edge of $G_{p\times k}$ is parallel
to the $z$-axis.  All $r$ of these graphs are drawn on the same set of
$kx$ vertices and the total number of edges in all $r$ of these graphs
is $3kxt - O(kr + x)$.  Projecting the resulting 3-d geometric graph
in the same manner as the proof of \thmref{LB_thm2} yields the desired
2-d geometric graph.
\end{proof}

\begin{figure}
  \begin{center}
    \includegraphics{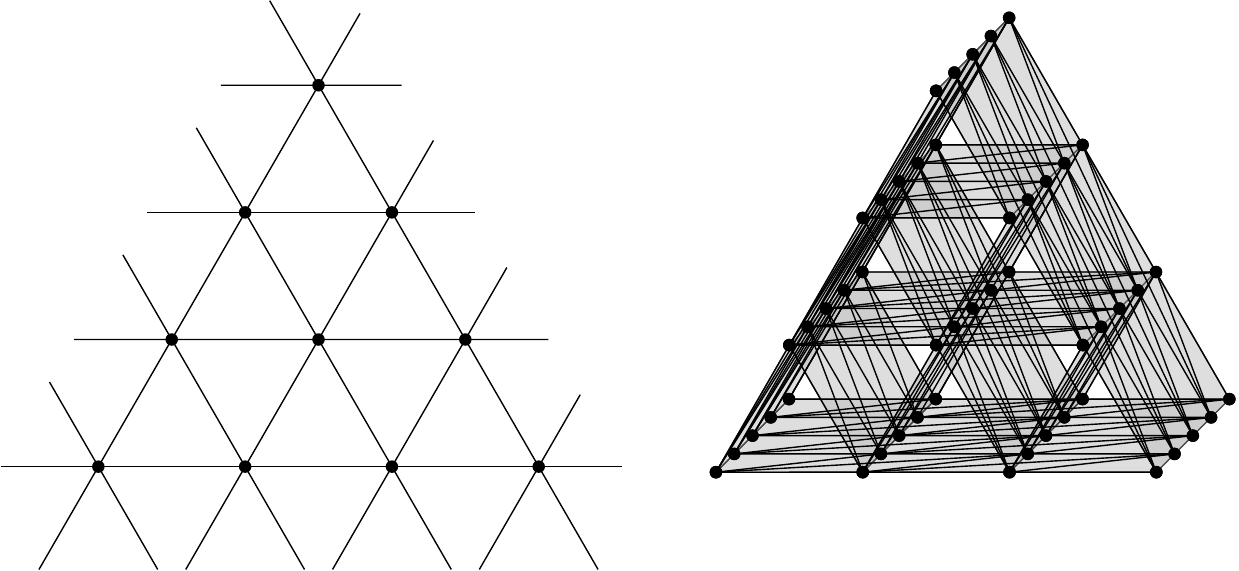}
  \end{center}
  \caption{An arrangement of $r=12$ lines having $x=10$ vertices of
           degree $2t=6$ yields a graph with $kx=50$ vertices and 
           $tkx - O(kr+x)$ edges.}
  \figlabel{convert}
\end{figure}

The next result applies \lemref{convert} to some standard grid
constructions:

\begin{thm}\label{thm:finite}
For any $\epsilon>0$, any $\alpha\in\{\pi/2,\pi/3,\pi/4,\pi/6\}$, and
all sufficiently large integers $n$, there exist $(\alpha-\epsilon)$AC
graphs that have $n$ vertices and $3(\pi/\alpha)n - O(n^{2/3})$ edges.
\end{thm}

\begin{proof}
Refer to \figref{lattices}.
As we have already seen, for $\alpha=\pi/2$ we use a set of $r$ horizontal
and vertical lines and apply \lemref{convert}.  For $\alpha=\pi/3$ we use
lines that support the triangular lattice.  For $\alpha=\pi/4$ we use
horizontal and vertical lines as well as lines of slope 1 and $-1$. For
$\alpha=\pi/6$, we use a further refinement of the triangular lattice.
\begin{figure}
  \begin{center}
    \includegraphics{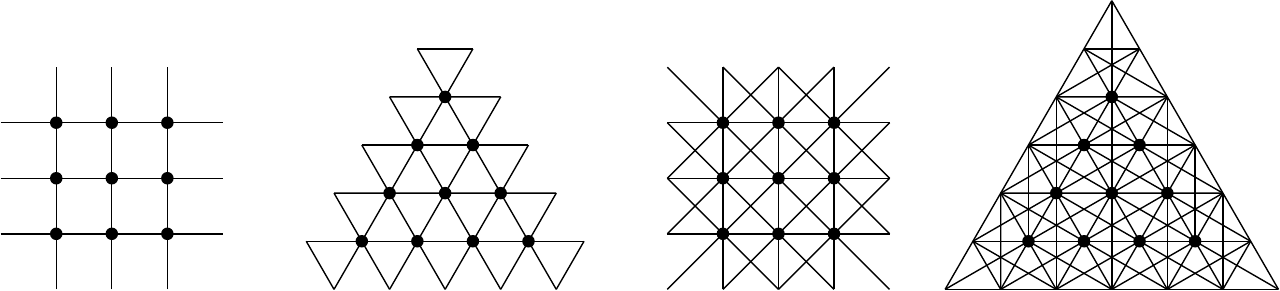}
  \end{center}
  \caption{The lattices used to prove lower bounds in \thmref{finite}.}
  \figlabel{lattices}
\end{figure}
\end{proof}

Finally, we prove a lower bound for any $\alpha$ of the form
$\pi/t-\epsilon$:
\begin{thm}
For any $\epsilon>0$, any integer $t\ge 2$, and all sufficiently large
integers $n$, there exist $(\pi/t-\epsilon)AC$ graphs that have $n$
vertices and $3tn - O(tn^{2/3}/\epsilon)$ edges.
\end{thm}

\begin{proof}
Set $\delta=\epsilon/2$.  We prove this result by showing the
existence of a set of $O(tr/\delta)$ lines satisfying the conditions of
\lemref{convert} with $\alpha=\pi/t-\delta$ and $x=r^2$.  We can then
apply \lemref{convert} with $k=r$, giving a $(\alpha = \pi/t-\epsilon)$AC
graph, with $n=r^3$ and $3tr^3 - O(tr^2/\epsilon + r^2) = 3tn -
O(tn^{2/3}/\epsilon)$ edges, as required.

We prove the existence of this set of lines as follows: We select a set
of $t$ lines whose slopes are rational, but very close to the directions
$\{i\pi/t:i\in\{0,\ldots,t-1\}\}$.  We then use $O(r/\epsilon)$
translates of these $t$ lines to cover each of the vertices of an $r\times
r$ grid by lines with each of the $t$ slopes.

The set of lines that we use will be such that all $r^2$ vertices of
the $r\times r$ grid $P=\{(i,j): i,j\in\{0,\ldots,r-1\}\}$ will have
$t$ lines passing through them.  We do this by first constructing a
\emph{$t$-frame} $F$ consisting of $t$ lines through the origin so that
any two lines of the $t$-frame meet at angles of at least $\pi/t-\delta$.
This $t$-frame also has the property that each line passes through a
grid point $(a,b)\neq(0,0)$ such that $|a|,|b|\le q$ for some integer $q\in
O(1/\delta)$. (See \figref{bigproof}.a.)

\begin{figure}
  \begin{center}
    \begin{tabular}{ccc}
    \includegraphics[width=1.7in]{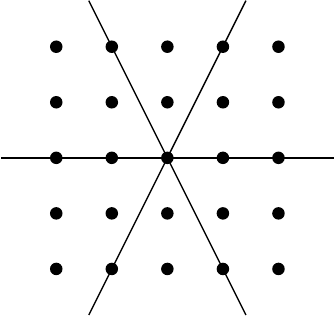}
    &
    \includegraphics[width=1.7in]{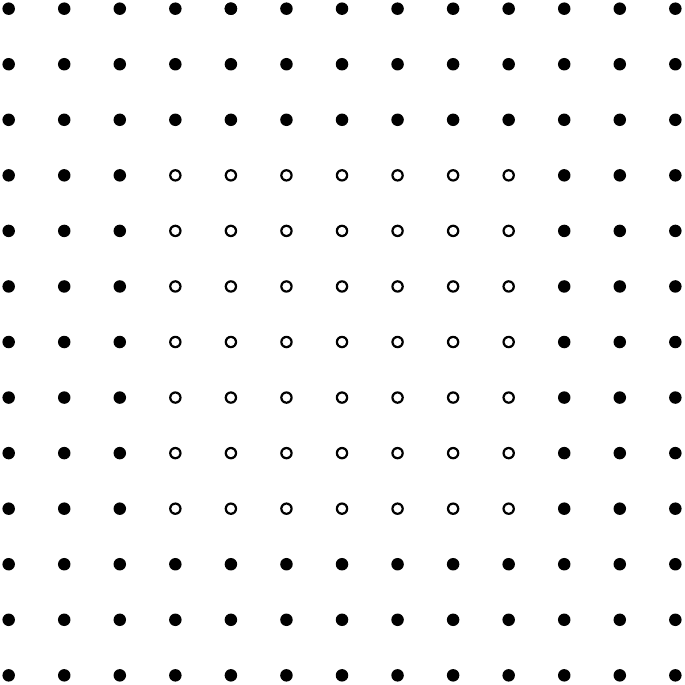}
    &
    \includegraphics[width=1.7in]{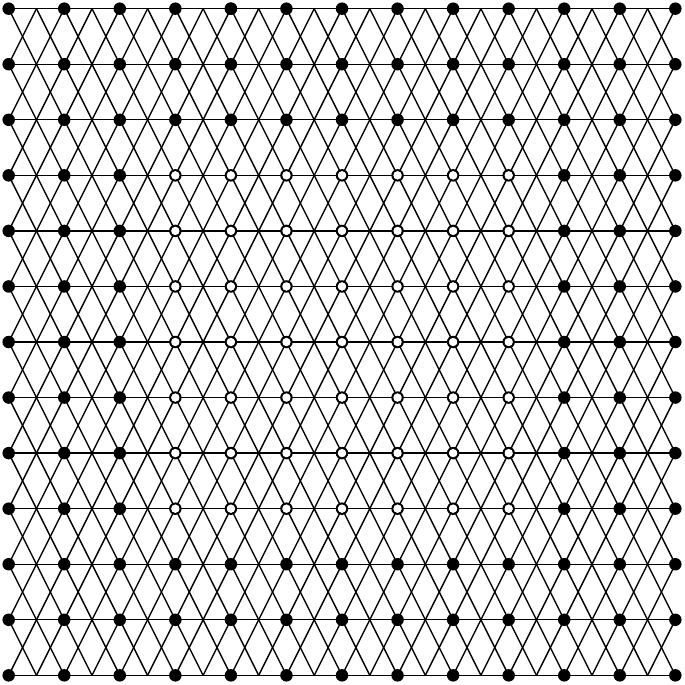} \\
    (a) & (b) & (c)
    \end{tabular}
  \end{center}
  \caption{(a)~A 3-frame, $F$, in which each line passes through a grid point
$(a,b)$ with $|a|,|b| \le 2$; 
           (b)~The set $P'$, for $k=2$; and 
           (c)~Placing $F$ at each point in $P'$ covers each point of
$P$ with 3 lines}
  \figlabel{bigproof}
\end{figure}

\newcommand{\floor}[1]{\lfloor #1 \rfloor}

Next, we make $O(r/\epsilon)$ translated copies of the frame $F$ by
translating $F$ to each of the points in the set
\begin{eqnarray*}
   P'  & = & \{(a,b) : a\in\{0,\ldots,r-1\},\, b\in\{0,\ldots,q\} \} \\
   & & {} \cup
   \{(a,b) : a\in\{0,\ldots,q\},\, b\in\{0,\ldots,r-1\} \} \\
   & & {} \cup 
   \{(a,b) : a\in\{r-1-q,\ldots,r-1\},\, b\in\{0,\ldots,r-1\} \} \\
   & & {} \cup
   \{(a,b) : a\in\{0,\ldots,r-1\},\, b\in\{r-1-q,\ldots,r-1\} \}
\end{eqnarray*}
(These are the points that are within distance $q$ of the boundary of
the grid $P$; see \figref{bigproof}.b.)  Denote the resulting set of
$O(tr/\epsilon)$ lines by $S$.  Notice that, for every point $(i,j)$ in
$P$, there are $t$ lines of $S$ that contain $(i,j)$.  This follows from
the fact that each line in the frame $F$ intersects all lattice points
$\{\lambda(a,b): \lambda\in \Z\}$ for some $a,b\in\Z$ with $|a|,|b|\le k$
(see \figref{bigproof}.c).  Thus, the set $S$ of lines, combined with
\lemref{convert} proves the theorem.  To complete the proof, all that
remains is to describe the frame $F$ used to construct $S$.

The frame $F$ consists of $t$ lines $L_0,\ldots,L_{t-1}$, all passing
through the origin, and such that the direction of $L_i$ is in $[i\pi/t
- \delta/2,i\pi/t+\delta/2]$.  Consider the \emph{ideal line} $L'_i$
that contains the origin and that has direction exactly $i\pi/t$.
Take a point $p$ on $L'_i$ whose distance from the origin is
$1/(\sqrt{2}\sin(\delta/2))$ (see \figref{angle}).  There is a point
$p'=(a,b)\in\Z^2$ whose distance from $p$ is at most $1/\sqrt{2}$.
The angle $\angle pOp'$ is at most $\delta/2$.  The distance from $p'$
to the origin is at most
\[
  1/(\sqrt{2}\sin(\delta/2))+1/\sqrt{2}
    \le 2\sqrt{2}\pi/\delta+1/\sqrt{2}
    = O(1/\delta) \enspace .
\]
Therefore $|a|,|b| \in O(1/\delta)$.  For our frame $F$, we take the line
$L_i$ that contains the origin and $(a,b)$.  This yields a set $F$ of lines
such that the angle between any two lines is at most $\delta$, and
completes the proof.
\end{proof}

\begin{figure}
  \begin{center}
    \includegraphics{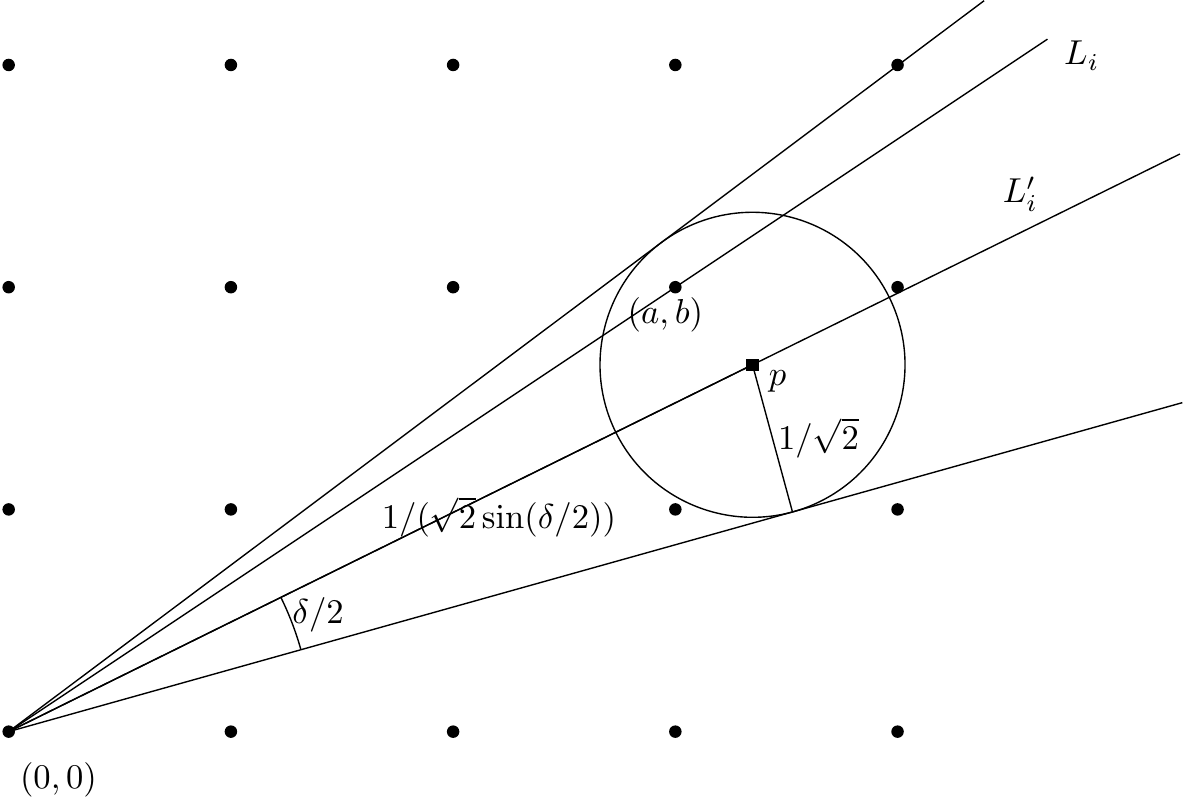}
  \end{center}
  \caption{For every line $L_i'$ through the origin with direction
$i\pi/t$, there exists a line $L_i$ through the origin and $(a,b)$ with
direction $d\pm\delta/2$ and with $|a|,|b|\in O(1/\delta)$}
  \figlabel{angle}
\end{figure}

\section{Charging Arguments}
\seclabel{charging}

In this section we derive upper bounds using charging arguments similar
to those used by Ackerman and Tardos \cite{at-mneqp-07} and Ackerman
\cite{a-mnetg-09}.  Let $G$ be an $\alpha$AC graph.  We denote by $G'$
the planar graph obtained by introducing a vertex at each point where
two or more edges in $G$ cross (thereby subdividing) edges of $G$.

For a face $f$ of $G'$, we denote by $|f|$ the length of the facial walk around $f$
so that, if we walk along an edge twice during the traversal, then it contributes twice to
$|f|$.  Let $v(f)$ denote the number of steps of this traversal during
which a vertex of $G$ (as opposed to a vertex introduced in $G'$) is
encountered.  For each face $f$ of $G'$ define the \emph{initial charge}
of~$f$ as
\[
    \ch(f) = |f| + v(f) - 4  \enspace .
\]
Ackerman and Tardos show, using two applications of Euler's formula, that
\[
    \sum_{f\in G'} \ch(f) = 4n-8 \enspace .
\]
We call a face $f$ of $G'$ a $k$-\emph{shape} if $v(f)=k$ and $f$ is a
\emph{shape}.  For example, a 2-pentagon is a face of $G'$ with $|f|=5$ and
$v(f)=2$.

As a warm-up, and introduction to charging arguments, we offer an alternate
proof to the upper bound presented by Didimo \etal\ in~\cite{del-dgrac-09}.

\begin{thm}
A RAC graph with $n \ge 4$ vertices has at most $4n-10$ edges.
\end{thm}
\begin{proof}
Let $G$ be a maximal RAC graph on $n$ vertices, and define $G'$ and $\ch$
as above.  We claim that, for every face $f$ of $G'$, $\ch(f)\ge v(f)/2$.
To see this, observe that the claim is certainly true if $|f| \ge 4$.  On
the other hand, if $|f|=3$ then, by the RAC property, $v(f) \ge 2$, so it
is also true in this case.
Therefore,
\begin{eqnarray*}
4n-8    & =  &  \sum_{f\in G'} \ch(f)  \ge  \sum_{f\in G'} v(f)/2 = \\
        & = &  \sum_{v\in G} \deg(v)/2  =  |E(G)| \enspace ,
\end{eqnarray*}
which proves that $E(G)\le 4n-8$.

To improve the above bound, observe that, since $G$ is maximal all vertices
on the outer face,~$f$, of $G'$ are vertices of $G$.  If $|f| \ge 4$ then
$\ch(f) \ge v(f)/2 + 2$, so in this case, proceeding as above, we have
\[
    4n-8-2 \ge |E(G)|
\]
and we are done.  Otherwise, the outer face of $G'$ is a 3-triangle and
$\ch(f) = v(f)/2 + 1/2$.
Consider the internal faces of $G'$ incident to the three edges of $f$.
Because $G$ is maximal, and $n\ge 4$, there must
be three such faces and each of these three faces, $f'$, has $v(f') \ge 2$.
Furthermore, at most one of these faces is a 2-triangle.\footnote{This is proven by a simple geometric argument that shows for
any triangle $f$, two right-angle triangles that are interior to $f$
and each share an edge with $f$ must overlap. See, e.g., the proof of
Theorem~1 in Reference~\cite{del-dgrac-09}.}
Consider each of the other two faces, $f'$.  If $|f'|=3$ then
\[
   \ch(f') = |f'| + v(f') - 4 = 2 = v(f')/2 + 1/2 \enspace ,
\]
since $v(f')=3$.
Otherwise, $|f'|\ge 4$, and we have 
\[
   \ch(f') = |f'| + v(f') - 4 \ge v(f') > v(f')/2 + 1/2 \enspace ,
\]
since $v(f')\ge 2$.
Therefore, we have
\[
    4n-8-3/2 \ge |E(G)|
\]
which, implies that $|E(G)| \le 4n-10$ since $|E(G)|$ is an integer.
\end{proof}

Next, we prove an upper bound for $\alpha > 2\pi/5$ that improves on the
$6.5n-20$ upper bound that follows from Ackerman and Tardos' bound on
quasiplanar graphs.

\begin{thm}\label{thm:six-n}
Let $G$ be an $\alpha$AC graph with $n$ vertices, for $\alpha > 2\pi/5$.
Then $G$ has at most $6n-12$ edges.
\end{thm}

\begin{proof}
We will redistribute the charge in the geometric graph $G'$ to obtain a new charge $\ch'$
such that $\ch'(f) \ge v(f)/3$ for every face $f$ of $G$.  In this way, we
get
\begin{eqnarray*}
4n-8    & =  &  \sum_{f\in G'} \ch(f)\\
        & =  &  \sum_{f\in G'} \ch'(f)\\
        & \ge & \sum_{f\in G'} v(f)/3\\
        & = &   \sum_{v\in G} \deg(v)/3\\
        & = & 2|E(G)|/3 \enspace ,
\end{eqnarray*}
which we rewrite to get $|E(G)| \le 6n-12$.

The charge $\ch'(f)$ is obtained as follows.  Let $f$ be any 1-triangle of
$G'$.  (Note that $\ch(f) = 0$.)   That is, $f$ is a triangle formed by two
edges $e_1$ and $e_2$ that meet at a vertex $x$ of $G$ and an edge $e$ that
crosses $e_1$ and $e_2$.
Imagine walking along the bisector of $e_1$ and $e_2$ (starting in the
interior of $f$) until reaching a face $f'$ such that $f'$ is not a
0-quadrilateral.  To see why such an $f'$ exists, observe that if we
encounter nothing but 0-quadrilaterals we will eventually reach a face that
contains an endpoint of $e_1$ or $e_2$ and is therefore not a
0-quadrilateral.

Adjust the charges at $f$ and $f'$ by subtracting $1/3$ from $\ch(f')$ and
adding $1/3$ to $\ch(f)$.  It is helpful to think of the charge as leaving
$f'$ through the last edge $e'$ traversed in the walk.  Note that neither
endpoint of $e'$ is a vertex of $G$.  This implies that for a face $f'$,
the amount of charge that leaves~$f'$ is at most
\begin{equation}
   \ell(f') \le \left\{
            \begin{array}{ll}
              |f'|         & \mbox{if $v(f')=0$} \\
              |f'| - v(f') - 1 & \mbox{otherwise.}
            \end{array}
          \right.
  \eqlabel{leaving}
\end{equation}
Let $\ch'$ be the charge obtained after performing this redistribution of
charge for every 1-triangle $f$.  We claim that $\ch'(f) \ge v(f)/3$ for every face $f$ of $G'$.  To
see this, we need only run through a few cases that can be verified using
\eqref{leaving} and the following observations:

\begin{enumerate}
\item If $|f|\ge 6$, then $\ell(f) \le |f|/3$, so $\ch'(f) \ge v(f) \ge v(f)/3$.

\item If $|f|=5$, then $v(f) \ge 1$ since, otherwise, $f$ has two edges on
its boundary that cross at an angle of less than or equal to $2\pi/5$.

\item If $|f|=4$, and $f$ is a 0-quadrilateral then $\ell(f)=0$, by construction.

\item If $|f|=3$, and $f$ is a 1-triangle then $\ch'(f)=1/3$, by construction.

\item If $|f|=3$ then $v(f)\ge 1$ since, otherwise, $f$ has two edges on
its boundary that cross at an angle less of at most $\pi/3 < 2\pi/5$.
\end{enumerate}
This completes the proof.
\end{proof}

\section{Notes}

Theorem~\ref{thm:six-n} appears to be true even for $\alpha > \pi/3$, but we have not
been able to prove it. The problem occurs because 0-pentagons can finish with
a charge of $-2/3$ or $-1/3$. (See \figref{pentagons}.)

\begin{figure}
  \begin{center}
    \includegraphics{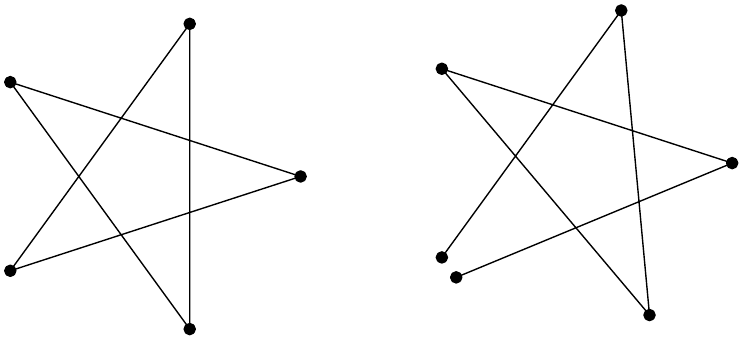}
  \end{center}
  \caption{Pentagrams lead to 0-pentagons with negative charge.}
  \figlabel{pentagons}
\end{figure}

A possible proof could look for extra charge near the vertices of the
penta\emph{gram} that created this pentagon, but it is easy to make gadgets
so that the faces surrounding those vertices have no extra charge.  Another
option is to look for extra charge near the vertices of the penta\emph{gon}
itself. Again, it is not too hard to to make them have no extra charge.
(See \figref{p2}.)

\begin{figure} [bth]
  \begin{center}
    \includegraphics{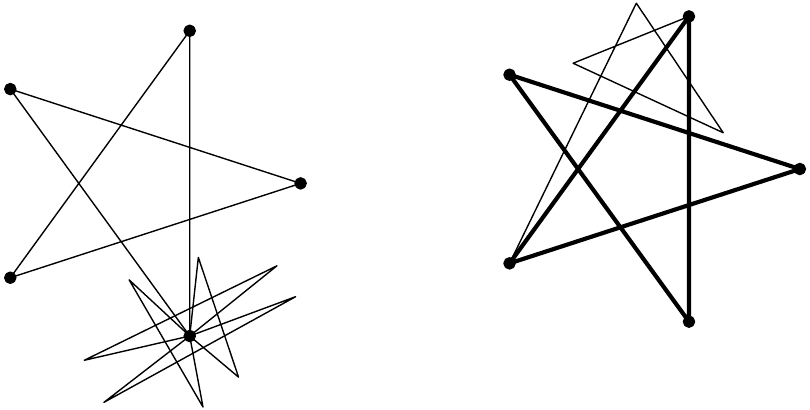}
  \end{center}
  \caption{Pentagrams can have 0 extra charge at their vertices and
           0 extra charge at the vertices of a pentagon.}
  \figlabel{p2}
\end{figure}

We also tried to follow the Ackerman-Tardos proof more closely. Namely, we
distribute the charge so that $\ch'(f)\ge v(f)/5$ and then prove that there
is leftover charge at the faces around each vertex.  For this to give a
bound of $6n$ we would need the extra charge at each vertex to be $8/5$.
Unfortunately, the limiting case in Ackerman-Tardos is $7/5$ and this is
realizable even with crossing angles arbitrarily close to $\pi/2$.
(See \figref{at-bound}.)

\begin{figure}
  \begin{center}
    \includegraphics{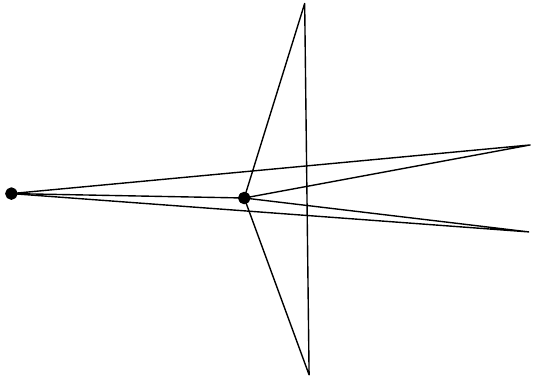}
  \end{center}
  \caption{The Ackerman-Tardos proof cannot even prove a bound of $6n$ for
          crossing angles of $\pi/2-\epsilon$.}
  \figlabel{at-bound}
\end{figure}

Finally, we can take a more global approach.  Discharging rules define a
directed graph among the faces (and possibly vertices) of $G'$.  An edge
$ab$ indicates that a charge of $x$ travels from $a$ to $b$, for some
number $x$ ($x=1/3$ in our argument).  The graph has to respect some flow
rules.  For example, in Theorem~\ref{thm:six-n} we have
\[
    \outd(a) - \ind(a) \le 3(|f| + 2v(f)/3 - 4) \enspace ,
\]
where $\ind$ and $\outd$ denote the in and out degree.  The goal would be
to define discharging paths recursively and then show that the recursion
terminates (i.e.~that the resulting graph is acyclic) and that the flow rule is satisfied.

\subsection*{Acknowledgements}
NICTA is funded by the Australian Government as represented by the
Department of Broadband, Communications and the Digital Economy and the
Australian Research Council through the ICT Centre of Excellence program.
Pat Morin's research was supported by NSERC, CFI, the Ontario Innovation Trust, NICTA, and the University of Sydney.

\bibliographystyle{plain}

\end{document}